\documentclass[letterpaper,titlepage,11pt]{article}

\pdfoutput=1

\usepackage{amssymb,amsmath,amsfonts}
\usepackage{epsfig}
\usepackage{ccaption}
\usepackage{graphicx}

\usepackage[
      colorlinks=true,
      linkcolor=blue,
      urlcolor=blue,
      filecolor=black,
      citecolor=red,
      pdfstartview=FitV,
      pdftitle={},
        pdfauthor={Roberto Emparan, Daniel Grumiller, Kentaro Tanabe},
        pdfsubject={},
        pdfkeywords={},
        pdfpagemode=None,
        bookmarksopen=true
      ]{hyperref}

\def\clock{{\count0=\time
           \divide\count0 60
           \ifnum\count0<10 0\fi\the\count0
           \multiply\count0 -60 \advance\count0 \time
           :\ifnum\count0<10 0\fi \the\count0
         }}
\newcommand{\timestamp}{{\small\vbox{\hbox{\tt\jobname.tex}
\hbox{\the\day/\the\month/\the\year, \clock}}}}

\newcommand{\ie}{{\it i.e.,\,}}
\newcommand{\eg}{{\it e.g.,\,}}

\newcommand{\lp}{\left(}
\newcommand{\rp}{\right)}

\newcommand{\beq}{\begin{equation}}
\newcommand{\eeq}{\end{equation}}
\newcommand{\bea}{\begin{eqnarray}}
\newcommand{\eea}{\end{eqnarray}}
\newcommand{\beqa}{\begin{eqnarray}}
\newcommand{\eeqa}{\end{eqnarray}}

\newcommand{\sR}{\mathsf{R}}

\begin{document}

\begin{titlepage}
\leftline{}
\vskip 1cm
\centerline{\LARGE \bf Large $D$ gravity and low $D$ strings 
} 
\vskip 1.6cm
\centerline{\bf Roberto Emparan$^{a,b}$, Daniel Grumiller$^{c}$, Kentaro Tanabe$^{b}$}
\vskip 0.5cm
\centerline{\sl $^{a}$Instituci\'o Catalana de Recerca i Estudis
Avan\c cats (ICREA)}
\centerline{\sl Passeig Llu\'{\i}s Companys 23, E-08010 Barcelona, Spain}
\smallskip
\centerline{\sl $^{b}$Departament de F{\'\i}sica Fonamental, Institut de
Ci\`encies del Cosmos,}
\centerline{\sl  Universitat de
Barcelona, Mart\'{\i} i Franqu\`es 1, E-08028 Barcelona, Spain}
\smallskip
\centerline{\sl $^{c}$Institute for Theoretical Physics, Vienna University of Technology}
\centerline{\sl Wiedner Hauptstrasse 8-10/136, A-1040 Vienna, Austria}
\vskip 0.5cm
\centerline{\small\tt emparan@ub.edu,\, grumil@hep.itp.tuwien.ac.at,\, ktanabe@ffn.ub.es}

\vskip 1.6cm
\centerline{\bf Abstract} \vskip 0.2cm \noindent

We point out that in the limit of large number of dimensions a wide class of non-extremal neutral black holes has a universal near horizon limit. The limiting geometry is the two-dimensional black hole of string theory with a two-dimensional target space. Its conformal symmetry explains properties of massless scalars found recently in the large $D$ limit. In analogy to the situation for NS fivebranes, the dynamics near the horizon does not decouple from the asymptotically flat region. We generalize the discussion to charged black $p$-branes. For black branes with string charges, the near horizon geometry is that of the three-dimensional
black strings of Horne and Horowitz. The analogies between the $\alpha'$ expansion in string theory and the large $D$ expansion in gravity suggest a possible effective string description of the large $D$ limit of black holes. We comment on applications to several subjects, in particular to the problem of critical collapse.

\end{titlepage}
\pagestyle{empty}
\small
\normalsize
\newpage
\pagestyle{plain}
\setcounter{page}{1}




\paragraph{I.} The study of General Relativity and its black holes simplifies drastically in the limit that the number $D$ of spacetime dimensions diverges \cite{Emparan:2013moa} (see also \cite{Asnin:2007rw}). The reason is that when $D$ is very large, the gravitational field is strongly localized within a region very close to the black hole horizon. Near a horizon of radius $r_0$ the potential develops a very large gradient, $\sim D/r_0$, with the result that the geometry further than a distance $\sim r_0/D$ from the horizon is essentially flat spacetime.

The appearance of two separate scales $r_0/D\ll r_0$ can be used to identify two different regions in the black hole geometry: a `far region', defined by $r-r_0\gg r_0/D$, and a `near-horizon region', where $r-r_0\ll r_0$. The dynamics in each of them is quite different: in the far region there are waves that propagate in flat spacetime; the near region contains the dynamics intrinsic to the black hole. These two sets of degrees of freedom interact in an `overlap region', $r_0/D\ll r-r_0\ll r_0$, common to both. Note that this two-region structure is a property of the geometry, and not the result of taking a long-wavelength approximation to field propagation in the black hole background.

In ref.~\cite{Emparan:2013moa} it has been argued that the existence of a far region in which the metric becomes exactly flat in the limit $D\to\infty$ is generic for very wide classes of black holes --- essentially, all black holes whose horizon length scales, as well as their asymptotic gravitational field, remain finite as $D\to\infty$. In this article we investigate the properties of the near-horizon region. We find that for many neutral black holes --- including all known vacuum and AdS non-extremal black holes whose size, again, remains finite in the limit $D\to\infty$ --- this region is universally described by a well-known geometry: the two-dimensional (2D) string-theory black hole of \cite{Mandal:1991tz,Elitzur:1991cb,Witten:1991yr}. The conformal symmetry of this solution is the explanation for the properties of the amplitudes for massless scalar fields in these backgrounds found in \cite{Emparan:2013moa}. Our results imply that the same symmetry is also present in the limit $D\to\infty$ near the horizon of many other neutral black holes, including rotating black holes and Anti-de~Sitter (AdS) black holes.

We also explore the large $D$ limit of a general class of dilatonic black $p$-branes, charged under a $(q+2)$-form field strength ($q\leq p$). We find that for non-dilatonic solutions with electric string charge the near-horizon region is the three-dimensional black string of Horne and Horowitz \cite{Horne:1991gn}. 

\paragraph{II.} Let us begin with the Schwarzschild--Tangherlini solution in $D=3+n$ dimensions
\beq\label{schwd}
ds^2=-f(r)\, dt^2+\frac{dr^2}{f(r)}+r^2 \,d\Omega_{n+1}^2\,, \qquad\qquad f(r)=1-\lp \frac{r_0}{r}\rp^{n}\,,
\eeq
where $d\Omega_{n+1}^2$ is the line-element of the round $(n+1)$-sphere,
and introduce the coordinate $\sR=(r/r_0)^n$, in terms of which
\beq
ds^2=-\frac{\sR-1}{\sR}\,dt^2+\frac{r_0^2}{n^2}\sR^{2/n}\,\frac{d\sR^2}{\sR(\sR-1)}+r_0^2\sR^{2/n} \,d\Omega_{n+1}^2\,.
\eeq
The near-horizon region at large $n$ is defined by $\ln \sR\ll n$, so we find
\beq
ds^2_\mathrm{nh}=-\frac{\sR-1}{\sR}\,dt^2+\frac{r_0^2}{n^2}\,\frac{d\sR^2}{\sR(\sR-1)}+r_0^2 \,d\Omega_{n+1}^2\,.
\eeq
We see that the size along the radial direction is very small when $n$ is large. As observed in \cite{Emparan:2013moa}, this region would be traversed very quickly, on a time $\sim r_0/n$, by freely falling observers. It is then convenient to rescale the time to $\hat t=n t/(2r_0)$. Furthermore, if we change $\sR=\cosh^2\!\rho$ the near-horizon metric becomes
\beq\label{nh1}
ds^2_\mathrm{nh}=\frac{4r_0^2}{n^2}\lp -\tanh^2\!\rho\, d\hat t^2+d\rho^2\rp +r_0^2 \,d\Omega_{n+1}^2\,.
\eeq
The part of the metric in parenthesis is the 2D string black hole of \cite{Mandal:1991tz,Elitzur:1991cb,Witten:1991yr}, which is the coset manifold $SL(2,\mathbb{R})/U(1)$. 
The appearance of this geometry in this context is actually expected from an observation made in \cite{Soda:1993xc,Grumiller:2002nm}. It is well known that the dimensional reduction on a sphere,  
\beq\label{redn}
ds^2=g_{\mu\nu}dx^\mu dx^\nu + r_0^2 e^{-4\Phi/(n+1)}d\Omega_{n+1}^2
\eeq
where $g_{\mu\nu}(x^\lambda)$ is a 2D metric and $\Phi(x^\lambda)$ a scalar field, of the Einstein-Hilbert action yields a specific 2D dilaton gravity action,
\beq
I=\frac{\Omega_{n+1}r_0^{n+1}}{16\pi G}\int d^2 x\sqrt{-g}\, e^{-2\Phi}
\lp R+\frac{4n}{n+1}(\nabla\Phi)^2 + \frac{n(n+1)}{r_0^2} e^{4\Phi/(n+1)}\rp\,.
\label{eq:angelinajolie}
\eeq
Maybe less well known \cite{Soda:1993xc,Grumiller:2002nm} is that in the limit $n\to\infty$ the action \eqref{eq:angelinajolie} is equivalent to the 2D string action
\beq
I=\frac{1}{16\pi G_2}\,\int d^2 x\sqrt{-g}\, e^{-2\Phi}
\lp R+4(\nabla\Phi)^2 + 4\lambda^2\rp
\label{eq:larged3}
\eeq
with
\beq
G_2 = \lim_{n\to\infty} \frac{G}{\Omega_{n+1}r_0^{n+1}}
\eeq
after we identify the cosmological constant parameter $\lambda=n/(2r_0)$. Notice that keeping $r_0/n$ finite amounts to keeping finite the Hawking--temperature, 
\beq\label{TH}
T_H=\frac{\lambda}{2\pi}\,
\eeq
of both the large $D$ Schwarzschild--Tangherlini and the 2D string black hole. The stretching of the time coordinate performed above has the effect of rescaling the Euclidean time circle to finite size. The large difference in the sizes of the 2D metric and the transverse $S^{n+1}$ in \eqref{nh1} makes the 2D gravitational constant very small if we keep fixed the 2D size $r_0/n$ and also fix $G$. This is particularly important in the quantum theory, as we discuss later.

The presence of the 2D string-theory black hole geometry in \eqref{nh1} implies that the amplitudes for propagation of waves in this background will realize the conformal symmetry $SL(2,\mathbb{R})$, thus providing a rationale for the results of \cite{Emparan:2013moa}. 

The Minkowski vacuum in the limit of large $n$ corresponds, after rescaling $t$ by a factor of $n$, to the linear dilaton vacuum of the 2D theory.

Other features of  fields in the background \eqref{schwd} also have a nice interpretation upon dimensional reduction on the sphere. The action for a minimal scalar $\Psi=\psi(t,r)Y^{(l)}_{n+1}(\Omega)$, where $Y^{(l)}_{n+1}(\Omega)$ are spherical harmonics on $S^{n+1}$, in the geometry \eqref{redn} is
\beqa
I[\Psi]&=&\frac12\int d^D x\sqrt{-g_{(D)}}	\lp\nabla_{(D)}\Psi\rp^2\notag\\
&=&
\frac{\Omega_{n+1}r_0^{n+1}}{2}\int d^2x\sqrt{-\hat g}\,e^{-2\Phi}\left[ (\hat\nabla\psi)^2+4e^{4\Phi/(n+1)}\frac{l}{n}\lp\frac{l}{n}+1\rp\psi^2\right]\notag\\
&\to&
\frac{\Omega_{n+1}r_0^{n+1}}{2}\int d^2x\sqrt{-\hat g}\,e^{-2\Phi}\left[ (\hat \nabla\psi)^2+4\frac{l}{n}\lp\frac{l}{n}+1\rp\psi^2\right]
\label{eq:larged4}
\eeqa
where we have rescaled the 2D metric $g_{\mu\nu}=\lambda^{-2}\hat g_{\mu\nu}$.
Consider first fields with $l\sim {\cal O}(n^0)$. These propagate in the 2D black hole geometry as massless, non-minimally coupled scalars. If their frequency in `far region time' $t$ is $\omega r_0\sim {\cal O}(n^0)$, then in `near-horizon time' $\hat t$ it is $\hat\omega=2\omega r_0/n \sim {\cal O}(n^{-1})$. These excitations encounter a dilaton barrier much higher than their energy and have vanishingly small amplitude for tunelling between the near and far regions, so that they can be said to decouple. Instead, waves of frequency $\hat\omega \sim {\cal O}(n^{0})$ have non-zero probability to penetrate or to pass above the barrier \cite{Emparan:2013moa}. 
Thus the large $D$ limit is not a decoupling limit. In this sense, these near-horizon geometries are similar to the near-horizon region of near-extremal NS fivebranes \cite{Maldacena:1997cg}.

Waves with large angular momentum $\hat l=2l/n={\cal O}(n^0)$ have an effective mass $\sim \hat l/r_0$. They must have frequency $\hat\omega >\hat l+1$ in order to escape to the asymptotic region \cite{Emparan:2013moa}. These excitations
probe scales much smaller than the radius of the ${n+1}$-sphere and effectively see the geometry
\beq\label{nh2}
ds^2_\mathrm{nh}=\frac{4r_0^2}{n^2}\lp -\tanh^2\rho\, d\hat t^2+d\rho^2 + d\mathbf{x}^2_{n+1}\rp\,.
\eeq

We can verify the presence of the same geometry in other neutral black holes. Consider the Myers-Perry solutions with (for simplicity) a single rotation \cite{Myers:1986un}. In terms of the coordinate $\sR$ introduced above, they take the form
\beqa
ds^2&=&-dt^2+\frac{1}{\sigma\sR}\lp dt +\alpha r_0 \sin^2\!\theta \,d\phi\rp^2 \notag\\
&&+
\sR^{2/n}r_0^2 \Biggl(
\frac{\sigma}{n^2\delta}\frac{d\sR^2}{\sR^2}+\sigma \,d\theta^2+\lp 1+\alpha^2\sR^{-2/n}\rp
\sin^2\!\theta\, d\phi^2+\cos^2\!\theta\, d\Omega_{n-1}^2\Biggr)
\eeqa
where
\beq
\sigma=1+\frac{\alpha^2\cos^2\!\theta}{\sR^{2/n}}\,,\qquad 
\delta =1+\frac{\alpha^2}{\sR^{2/n}}-\frac1{\sR}\,.
\eeq
The usual rotation parameter is $a=\alpha r_0$. We see again that the metric in radial directions is small. So we consider waves of frequency $\omega = {\cal O}(n)$, which are the ones that can probe the full geometry. The metric that a partial wave sees will be different depending on the value of the component $\l_\phi$ of angular momentum along the rotation direction $\phi$. 
When the wave has small impact parameter $l_\phi/\omega$ in the plane of rotation, as is the case when $l_\phi=\mathcal{O}(n^0)$, it probes the region near the pole at $\theta=0$. At larger impact parameters, with $l_\phi= \mathcal{O}(n)$, it probes the region around a finite angle $\theta_0$. Denoting $v=\alpha\sin\theta_0/\sqrt{1+\alpha^2}$, the appropriate rescalings are
\beq
\hat t=\frac{n}{2r_0}\frac{t}{\sqrt{1-v^2}}\,,\qquad \hat y=\frac{n(1+\alpha^2)}{2\alpha}\frac{v}{\sqrt{1-v^2}}\,\phi\,,\qquad \hat\theta=\frac{n\sqrt{1+\alpha^2}}{2}\lp\theta_0-\theta\rp\,.
\eeq
If we set 
$\sR=(1+\alpha^2)^{-1}\cosh^2\rho$ we find
\beqa
ds^2 &=& \frac{4r_0^2}{n^2}(1-v^2) \,\bigg(-d\hat t^2+d\hat y^2+\frac{\lp d\hat t+v\,d\hat y\rp^2}{(1-v^2)\cosh^2\rho} + d\rho^2\notag\\
&&+ d\hat\theta^2+\frac{n^2}{4(1-v^2)}\cos^2\theta\, d\Omega^2_{n-1}\bigg)\,.
\eeqa
The $(\hat t,\hat y,\rho)$ part of the metric is the result of adding a line $\hat y$ to the 2D string black hole and performing a boost of velocity $v$ along $\hat y$. So locally this geometry is equivalent to the 2D string black hole. For angular momenta that are ${\cal O}(n^0)$ the wave probes the pole region where $v=0$ and the geometry is the product of the static 2D string black hole times $\mathbb{R}^2\times S^{n-1}$. When the impact parameter is maximum, so that $\theta_0=\pi/2$, the wave is strongly localized in the angular directions transverse to the rotation direction and the $(\hat\theta,\Omega_{n-1})$ part of the metric is actually $\mathbb{R}^{n}$. 

These conclusions generalize to the case in which the black hole has several non-zero spins: for ${\cal O}(n)$ frequencies and angular momenta along a direction in which the black hole rotates,  we find the 2D string black hole with a boost along the rotation direction. For small, $\mathcal{O}(n^{-1})$, impact parameters  we recover the static 2D black hole.

When the number of non-zero spins grows like $n/2$ the situation requires some slight modifications. As an illustrative case we take a black hole in odd dimensions with all possible rotation parameters turned on, and all equal to each other, $a_i=\alpha r_0$. The radial coordinate of the 2D black hole, in terms of the Boyer-Lindquist radius $r$ \cite{Myers:1986un}, is
\beq
\cosh^2\rho =\lp \frac{r}{r_H}\rp^{n(1-\alpha^2)}\,.
\eeq
Here the horizon radius $r_H$ is
\beq
r_H=r_0\sqrt{1-\alpha^2}\lp 1+\frac1n\frac{\ln (1-\alpha^2)}{1-\alpha^2}+{\cal O}(n^{-2})\rp
\eeq
and we assume that $\alpha<1$ remains fixed as $n\to\infty$. Then, rescaling $t$ and the angles appropriately as before, we recover the boosted 2D string black hole. In the cases where all the spins of the black hole are turned on, the solution admits an extremal limit. This would correspond in this example to $\alpha\to 1$, for which the above expansion breaks down. Extremal rotating black holes have zero temperature and we expect their near-horizon geometry to be locally inequivalent to the 2D string black hole. We shall not discuss in this article the detailed analysis required to study their near-horizon geometry. 

Let us now consider non-rotating AdS black holes. Their metric is of the same form as \eqref{schwd} but with
\beq
f(r)=1-\lp\frac{r_0}{r}\rp^{n}+\frac{r^2}{L^2}\,.
\eeq
Thus, when we take the large $n$ limit keeping the coordinate $\sR=(r/r_0)^n$ finite we find
\beq
ds^2=- \frac{\sR/\sR_0-1}{\sR}dt^2+\frac{r_0^2}{n^2}\frac{d\sR^2}{\sR(\sR/\sR_0-1)}
+r_0^2 d\Omega_{n+1}^2\,,
\label{eq:larged1}
\eeq
where 
\beq
\sR_0=\frac{L^2}{r_0^2+L^2}\,.
\label{eq:larged2}
\eeq
The only change relative to \eqref{nh1} is a shift in the location of the horizon to $\sR=\sR_0<1$. It can be absorbed in a rescaling of the time coordinate and angles as we have done in the rotating case. It simply amounts to a change in the mass of the 2D string black hole. The result also extends to black branes (planar black holes) in AdS. When rotation is added, the results parallel the ones we have obtained for vacuum black holes.
Our result \eqref{eq:larged1}-\eqref{eq:larged2} suggests that the flat space limit (sending the AdS radius to infinity, $L\to\infty$) of AdS holography could be particularly simple in the large $D$ limit. It would be interesting to verify this and to consider also the large $D$ limit on the field theory side.

We conjecture that the appearance of the 2D string black hole is a universal feature of the large $D$ limit of non-extremal neutral black holes, at least when the limit is taken in such a manner that the length scales of the horizon remain finite in the limit. Eq.~\eqref{nh2} is the geometry perceived by waves of  frequencies and angular momenta $\omega, l \sim {\cal O}(D)$.

We now examine black branes and include the effects of charge. We consider a large class of them, corresponding to dilatonic black $p$-branes with electric $q$-brane charge which are solutions of the theories
\beq\label{theory} 
I=\frac{1}{16\pi G} \int d^Dx \sqrt{-g}\left(
R-2(\nabla\varphi)^2-\frac{1}{2(q+2)!} e^{-2a\varphi}H_{[q+2]}^2 \right)\,.
\eeq 
We refer to appendix~A.2 of \cite{Caldarelli:2010xz} for details of the solutions. For a $p$-brane we denote $D=n+p+3$. The brane carries $q$-brane charge, with $q\leq p$, and the dilaton coupling is $a$. The charge is parametrized in terms of a charge-boost velocity $u$, so the more conventional charge rapidity $\beta$ is $\tanh\beta=u$. The parameter $u$  varies between $0$ for neutral branes and $u=1$ for extremal ones, and it is kept finite as $D\to\infty$, and so are $p$ and $q$, too. In this limit, the near-horizon geometries become (we set $r_0=1$ for simplicity)
\beq\label{dilpqbranes}
ds^2=\lp 1-\frac{u^2}{\sR}\rp^N \lp -\frac{\sR-1}{\sR-u^2}dt^2 + d\mathbf{y}^2_q\rp
+\frac1{n^2}\frac{d\sR^2}{(\sR-u^2)(\sR-1)}+d\Omega_{n+1}^2+d\mathbf{z}^2_{p-q}\,,
\eeq
where
\beq
N=\frac{4}{2(q+1)+a^2}\,.
\eeq
All the dependence of metric functions on $p$, $q$ and $a$ has been reduced to  a single parameter $N$. This may be an indication of universality classes of near-horizon solutions labeled by the values of $N$ and $u$. 

The presence of charge deforms the near horizon geometry away from the 2D string black hole to other black holes of dimensionally-reduced (multi-)dilaton theories. Let us consider the case $q=1$, $a=0$, \ie\ $N=1$. After rescaling the $t$ and $y$ coordinates by a factor of $n$ we obtain
\beq
n^2 ds^2=-\lp 1-\frac{1}{\sR}\rp d\hat t^2+\lp 1-\frac{u^2}{\sR}\rp d\hat y^2
+\frac{d\sR^2}{(\sR-u^2)(\sR-1)}+n^2 d\Omega_{n+1}^2+n^2 d\mathbf{z}^2_{p-1}\,.
\eeq
The sector $(\hat t, \hat y, \sR)$ in this metric is the geometry of the three-dimensional black strings of \cite{Horne:1991gn}, which are a solution of three-dimensional string theory with a known exact conformal field theory. This geometry arises whenever $q=1$ and $a=0$, independently of $p$. Its appearance should not be a surprise when we consider that it can be obtained by adding a line $y$ to the 2D string black hole followed by a boost and T-duality along $y$ \cite{Horne:1991cn}.

These  geometries arise in the near-horizon region of non-extremal black holes, with no requirement that the solutions be any close to extremality. For extremal solutions the variety of near-horizon geometries can be expected to be much larger. We can easily find some examples. For instance, the large $D$ limit 
of the Reissner-Nordstrom black holes is obtained setting $a=0=p=q$ in the solutions above, and then the extremal limit is reached when $u\to 1$. We find
\beqa
n^2 ds^2&=&-\lp 1-\sR^{-1}\rp^2 d\hat t^2+\frac{d\sR^2}{(\sR-1)^2}+n^2d\Omega_{n+1}^2\notag\\
&=&
-\frac{1}{(1+e^{-\rho})^2}\,d\hat t^2+d\rho^2+n^2d\Omega_{n+1}^2\,,
\eeqa
where in the last line we have changed to $\rho=\ln (\sR-1)$.
The $(\hat t, \rho)$ part of the geometry approaches 2D flat spacetime in the asymptotic region $\rho\to \infty$, while near the horizon at $\rho\to-\infty$ it becomes the infinite throat of the AdS$_2$ spacetime. So the large $D$ limit results in the replacement of the non-extremal horizon of the 2D string black hole with an extremal throat characteristic of the solution at finite $D$. We expect this type of replacement to be a general feature of large $D$ extremal charged solutions.

\paragraph{III.} The emergence of a 2D conformal symmetry offers, at the very least, the prospect of a significant degree of control over the classical theory of very wide classes of neutral, non-extremal large $D$ black holes. Even more tantalizing is the appearance of low-dimensional string geometries near the horizon of large $D$ black holes. Regarding $D$ as a parameter that can be made large is essential for the stringy interpretation. The connection to the 2D string black hole instructs us to identify
\beq
\sqrt{\alpha'}\sim \frac{r_0}{D}\,,
\eeq
so the large $D$ expansion corresponds to the $\alpha'$ expansion in string theory. The near-horizon geometries are `stringy geometries', of size $\sim r_0/D$. 
Note also that the Bekenstein--Hawking entropy $S_{BH}$ of Schwarzschild--Tangherlini black holes, as a function of the mass $M$, is
\beq
S_{BH} \propto M^{1+\frac1{D-3}}\,,
\eeq
and therefore in the large $D$ limit asymptotes to the Hagedorn behavior expected from string theory, $S_{BH} \propto M$ \cite{Emparan:2013moa}. Moreover, the black hole temperature  \eqref{TH} corresponds to the string scale $T_H\sim 1/\sqrt{\alpha'}$. This may be an indication of a string-like nature of the excitations near the horizon. 

In view of these observations, the two-region picture of large $D$ black hole spacetimes would have the following interpretation: in the far region we keep $r_0$ fixed, so when $D\to\infty$ we have $\alpha'\to 0$ and the excitations that remain in this region correspond to massless gravitons propagating in flat spacetime. In the near region we keep $r_0/D$ finite, and we obtain a string-scale geometry with string excitations. 
We note, however, that $\alpha'$ corrections to the 2D black hole are known \cite{Dijkgraaf:1991ba}, and they do not coincide with the $1/D$ corrections to the near-horizon geometry. Probably we should not be surprised by these discrepancies: since full quantum string theory cannot be formulated consistently in these large $D$ spacetimes, presumably any strings in the near-horizon region should be effective strings and not fundamental ones. 

Nevertheless, it would be remarkable to find, similarly to what happens in the large $N$ limit of Yang-Mills theories, that an effective string theory emerges in the large $D$ limit of gravity --- in this case, one has to look for the strings near a black hole horizon. If indeed this occurs, the symmetry $SO(D-2)$ of the angular sphere will likely play a role. More work is needed to put these ideas on a firmer ground and, if they are correct, identify the kind of effective string theories that can arise in this context.

The quantum theory may also be constrained by the near-horizon conformal symmetry. The strength of quantum gravitational effects can be controlled by suitably choosing how the Planck length scales with $D$ as the number of dimensions increases. Notice that the `effective string length' $r_0/D$ is a purely classical length scale (as it is in string theory) which can be arbitrarily separated from the quantum Planck scale, \eg\ it is possible to have large $\alpha'$ effects but small quantum gravity effects.

It should be interesting to study the evaporation of these black holes through quantum Hawking radiation. If the Planck length is chosen to not scale with $D$, then 
there is (at least) one significant difference with the situation for near-extremal NS five branes, which decay by the slow leakage of radiation of energy $\sim T_H$ that reaches the asymptotic region of the 2D black hole. For large $D$ black holes (and with a $D$-independent Planck scale) the typical energy of Hawking quanta is much larger, $\omega \sim D T_H\sim D^2/r_0$ \cite{Hod:2011zzb,Emparan:2013moa}. The wavelength of these quanta is so small that they do not distinguish between the near and far regions, and they accelerate enormously the decay rate of the black holes. However, if the Planck length $(G\hbar)^{1/(D-2)}$ is made to shrink like $1/D$ then typical Hawking quanta will have string-like energies $\sim D/r_0$ (analogous to the 't~Hooft large-$N$ limit in which the gauge coupling $g_{YM}^2$ is made to shrink like $1/N$). This may give a better chance of controlling the evaporation process by using the near-horizon 2D conformal symmetry and, possibly, an effective string description. 

Finally, from a purely pragmatic viewpoint, the large $D$ limit can lead to considerable simplifications \cite{Emparan:2013moa}, and the effective 2D formulation discussed in the present work can be a useful tool. As an example let us consider critical collapse {\`a} la Choptuik \cite{Choptuik:1992jv}. So far, analytic derivations of the critical exponent $\gamma$ are generally not possible. One relevant exception is the derivation in \cite{Strominger:1993tt} of the value $\gamma=\tfrac12$ in the RST model. This theory differs from the 2D model \eqref{eq:larged3}, \eqref{eq:larged4} in three aspects: it takes into account semi-classical corrections, it has a large number of scalar fields instead of just one, and the scalar fields do not couple to the dilaton field in 2D. Nevertheless, it seems plausible to us that their result for the critical exponent is the correct one in the large $D$ limit, since the geometric part of the RST model coincides classically with the action \eqref{eq:larged3}. There is also numerical evidence at $D\sim{\cal O}(10)$ that backs up our conjecture \cite{Sorkin:2005vz,Bland:2005kk}. It would be gratifying to derive our conjectured result for the critical exponent, $\gamma=\tfrac12$, analytically and to probe numerically the large $D$ regime. We believe that both avenues are accessible.

\section*{Acknowledgments}

RE was partially supported by MEC FPA2010-20807-C02-02, AGAUR 2009-SGR-168 and CPAN CSD2007-00042 Consolider-Ingenio 2010. 
DG was supported by the START project Y 435-N16 of the Austrian Science Fund (FWF).
KT was supported by a grant for research abroad by JSPS.


\end{document}